\documentclass[twocolumn,showpacs,aps,prl,superscriptaddress,floatfix]{revtex4}

\usepackage{graphicx}
\usepackage{epsfig}
\usepackage{amsmath}
\usepackage{amssymb}
\usepackage{hhline}
\usepackage{multirow}
\usepackage{fancyheadings}
\usepackage{ifthen}

\RequirePackage{xspace}

\RequirePackage{xspace}

\usepackage{relsize}
\def\babar{\mbox{\slshape B\kern-0.1em{\smaller A}\kern-0.1em
    B\kern-0.1em{\smaller A\kern-0.2em R}}}

\def\nub        {\ensuremath{\overline{\nu}}\xspace}

\def\nub        {\ensuremath{\overline{\nu}}\xspace}

\def\ssbar {\ensuremath{s\overline s}\xspace}

\def\piz   {\ensuremath{\pi^0}\xspace}

\def\Kbar  {\kern 0.2em\overline{\kern -0.2em K}{}\xspace}

\def\Kz    {\ensuremath{K^0}\xspace}
\def\Kzb   {\ensuremath{\Kbar^0}\xspace}
\def\KzKzb {\ensuremath{\Kz \kern -0.16em \Kzb}\xspace}
\def\Kp    {\ensuremath{K^+}\xspace}
\def\Km    {\ensuremath{K^-}\xspace}

\def\KpKm  {\ensuremath{\Kp \kern -0.16em \Km}\xspace}
\def\KS    {\ensuremath{K^0_{\scriptscriptstyle S}}\xspace} 
\def\KL    {\ensuremath{K^0_{\scriptscriptstyle L}}\xspace}

\def\Dbar    {\kern 0.2em\overline{\kern -0.2em D}{}\xspace}
\def\Db      {\ensuremath{\Dbar}\xspace}
\def\Dz      {\ensuremath{D^0}\xspace}
\def\Dzb     {\ensuremath{\Dbar^0}\xspace}
\def\DzDzb   {\ensuremath{\Dz {\kern -0.16em \Dzb}}\xspace}
\def\Dp      {\ensuremath{D^+}\xspace}
\def\Dm      {\ensuremath{D^-}\xspace}

\def\DpDm    {\ensuremath{\Dp {\kern -0.16em \Dm}}\xspace}

\def\Dstarzb {\ensuremath{\Dbar^{*0}}\xspace}
\def\Dstarp  {\ensuremath{D^{*+}}\xspace}

\def\B       {\ensuremath{B}\xspace}
\def\Bbar    {\kern 0.18em\overline{\kern -0.18em B}{}\xspace}
\def\Bb      {\ensuremath{\Bbar}\xspace}
\def\BB      {\ensuremath{B\Bbar}\xspace} 
\def\Bz      {\ensuremath{B^0}\xspace}
\def\Bzb     {\ensuremath{\Bbar^0}\xspace}
\def\BzBzb   {\ensuremath{\Bz {\kern -0.16em \Bzb}}\xspace}
\def\Bu      {\ensuremath{B^+}\xspace}
\def\Bub     {\ensuremath{B^-}\xspace}

\def\Bm      {\ensuremath{\Bub}\xspace}

\def\BpBm    {\ensuremath{\Bu {\kern -0.16em \Bub}}\xspace}

\mathchardef\Upsilon="7107
\def\Y#1S{\ensuremath{\Upsilon{(#1S)}}\xspace}

\def\FourS {\Y4S}

\mathchardef\Deltares="7101
\mathchardef\Xi="7104
\mathchardef\Lambda="7103
\mathchardef\Sigma="7106
\mathchardef\Omega="710A

\def\Deltabar{\kern 0.25em\overline{\kern -0.25em \Deltares}{}\xspace}
\def\Lbar{\kern 0.2em\overline{\kern -0.2em\Lambda\kern 0.05em}\kern-0.05em{}\xspace}
\def\Sigbar{\kern 0.2em\overline{\kern -0.2em \Sigma}{}\xspace}
\def\Xibar{\kern 0.2em\overline{\kern -0.2em \Xi}{}\xspace}
\def\Obar{\kern 0.2em\overline{\kern -0.2em \Omega}{}\xspace}
\def\Nbar{\kern 0.2em\overline{\kern -0.2em N}{}\xspace}
\def\Xb{\kern 0.2em\overline{\kern -0.2em X}{}\xspace}

\def\BR         {{\ensuremath{\cal B}\xspace}}

\def\mes        {\mbox{$m_{\rm ES}$}\xspace}

\newcommand{\tev}{\ensuremath{\mathrm{\,Te\kern -0.1em V}}\xspace}
\newcommand{\gev}{\ensuremath{\mathrm{\,Ge\kern -0.1em V}}\xspace}
\newcommand{\mev}{\ensuremath{\mathrm{\,Me\kern -0.1em V}}\xspace}
\newcommand{\kev}{\ensuremath{\mathrm{\,ke\kern -0.1em V}}\xspace}
\newcommand{\ev}{\ensuremath{\mathrm{\,e\kern -0.1em V}}\xspace}
\newcommand{\gevc}{\ensuremath{{\mathrm{\,Ge\kern -0.1em V\!/}c}}\xspace}
\newcommand{\mevc}{\ensuremath{{\mathrm{\,Me\kern -0.1em V\!/}c}}\xspace}
\newcommand{\gevcc}{\ensuremath{{\mathrm{\,Ge\kern -0.1em V\!/}c^2}}\xspace}
\newcommand{\mevcc}{\ensuremath{{\mathrm{\,Me\kern -0.1em V\!/}c^2}}\xspace}

\def\mus  {\ensuremath{\rm \,\mus}\xspace}

\def\ps   {\ensuremath{\rm \,ps}\xspace}

\def\mus        {\ensuremath{\,\mu{\rm s}}\xspace}    
\def\ps         {\ensuremath{{\rm \,ps}}\xspace}  

\def\ra                 {\ensuremath{\rightarrow}\xspace}
\def\to                 {\ensuremath{\rightarrow}\xspace}

\def\pep2{PEP-II}

\def\gsim{{~\raise.15em\hbox{$>$}\kern-.85em
          \lower.35em\hbox{$\sim$}~}\xspace}
\def\lsim{{~\raise.15em\hbox{$<$}\kern-.85em
          \lower.35em\hbox{$\sim$}~}\xspace}

\def\CP                {\ensuremath{C\!P}\xspace}

\def\Vub  {\ensuremath{|V_{ub}|}\xspace}

\xspace

\def\geant      {\mbox{\tt GEANT}\xspace}

\def\jetset     {\mbox{\tt Jetset}\xspace}

\def\B      {\ensuremath{B}\hbox{ }}

\newcommand {\Bxlnu}{\ensuremath{\Bb \rightarrow X \ell \bar{\nu}}}
\newcommand {\Bxclnu}{\ensuremath{\Bb \rightarrow X_c \ell \bar{\nu}}}
\newcommand {\Bxulnu}{\ensuremath{\Bb \rightarrow X_u \ell \bar{\nu}}}
\newcommand {\Bzxulnu}{\ensuremath{\Bzb \rightarrow X_u \ell \bar{\nu}}}
\newcommand {\Bpxulnu}{\ensuremath{\Bm \rightarrow X_u \ell \bar{\nu}}}
\newcommand {\Bpxlnu}{\ensuremath{\Bm \rightarrow X \ell \bar{\nu}}}

\newcommand {\Bzxlnu}{\ensuremath{\Bzb \rightarrow X \ell \bar{\nu}}}

\newcommand {\rusl}{\ensuremath{R_{u}}}
\newcommand {\mX}{\ensuremath{m_{X}}}
\newcommand {\mmiss}{\ensuremath{m_{miss}^2}}

\newcommand {\breco}{\ensuremath{B_{reco}}}

\newcommand {\D}{\ensuremath{D}}
\newcommand {\lbar}{\ensuremath{\overline{\Lambda}}}
\newcommand {\lone}{\ensuremath{\lambda_1}}
\newcommand{\beq}{\begin{equation}}
\newcommand{\beqa}{\begin{eqnarray}}
\newcommand{\beqn}{\begin{eqnarray}}
\newcommand{\eeq}{\end{equation}}
\newcommand{\eeqa}{\end{eqnarray}}
\newcommand{\eeqn}{\end{eqnarray}}
\makeatletter
\def\slash#1{{\mathpalette\c@ncel{#1}}} 
\makeatother

\def\allslsub{29982 \pm  233} 
 
\def\allnu{  175 \pm   21} 
\def\allbgc{   90 \pm    5} 
 
\def\allepsu{0.326} 
\def\allepsmx{0.770} 
\def\allbrbr{2.06 \pm 0.25}

\def\bchslsub{19080 \pm  191} 
 
\def\bchnu{  100 \pm   16} 
\def\bchbgc{   67 \pm    4}

\def\bchbrbr{1.82 \pm 0.30}

\def\bneslsub{10862 \pm  133} 
 
\def\bnenu{   76 \pm   15} 
\def\bnebgc{   22 \pm    3}

\def\bnebrbr{2.53 \pm 0.50}

\def\muslsub{12622 \pm  157} 
 
\def\munu{   73 \pm   15} 
\def\mubgc{   41 \pm    4}

\def\mubrbr{1.83 \pm 0.37}

\def\eleslsub{17320 \pm  173} 
 
\def\elenu{  101 \pm   15} 
\def\elebgc{   46 \pm    3}

\def\elebrbr{2.27 \pm 0.34}

\def\sboneslsub{ 4187 \pm   68} 
 
\def\sbonenu{   20 \pm    7} 
\def\sbonebgc{   12 \pm    1}

\def\sbonebrbr{1.68 \pm 0.57}

\def\sbtwoslsub{12373 \pm  141} 
 
\def\sbtwonu{   68 \pm   13} 
\def\sbtwobgc{   41 \pm    3}

\def\sbtwobrbr{1.94 \pm 0.37}

\def\sbthreeslsub{13144 \pm  170} 
 
\def\sbthreenu{   86 \pm   15} 
\def\sbthreebgc{   34 \pm    3}

\def\sbthreebrbr{2.31 \pm 0.41}

\def\lomxslsub{29982 \pm  233} 
 
\def\lomxnu{  143 \pm   18} 
\def\lomxbgc{   54 \pm    3}

\def\lomxbrbr{1.89 \pm 0.24}

\def\himxslsub{29982 \pm  233} 
 
\def\himxnu{  214 \pm   26} 
\def\himxbgc{  145 \pm    9}

\def\himxbrbr{2.35 \pm 0.28}

\def\allbrbrVal{2.06}
\def\allbrbrEcstat{0.25}
\def\allbrbrEsyst{0.23}
\def\allbrbrEthHi{0.36}

\def\allbr{ 2.24}
\def\allbrE{ 0.27}
\def\allbrEsyst{ 0.26}
\def\allbrEthHi{ 0.39}

\def\allvub{ 4.62}
\def\allvubE{ 0.28}
\def\allvubEsyst{ 0.27}
\def\allvubEthHi{ 0.40}

\def\allvubEtheo{ 0.26}

\renewcommand{\allepsu}{34.2}
\renewcommand{\allepsmx}{73.3}

\def\isprl{1}

\long\def\inst#1{\par\nobreak\kern 4pt\nobreak
    {\it #1}\par\vskip 10pt plus 3pt minus 3pt}

\def\varia1c-a{666}

\begin{document}

\title[Short Title] {Measurement   of   the    Inclusive   Charmless
Semileptonic Branching Ratio \\ of \mathversion{bold}$B$ Mesons and
Determination of \mathversion{bold}$|V_{ub}|$}

%
\author{B.~Aubert}
\author{R.~Barate}
\author{D.~Boutigny}
\author{J.-M.~Gaillard}
\author{A.~Hicheur}
\author{Y.~Karyotakis}
\author{J.~P.~Lees}
\author{P.~Robbe}
\author{V.~Tisserand}
\author{A.~Zghiche}
\affiliation{Laboratoire de Physique des Particules, F-74941 Annecy-le-Vieux, France }
% ...
\collaboration{The \babar\ Collaboration}
\noaffiliation

\begin{abstract}
We  report  a measurement  of  the  inclusive charmless  semileptonic
branching fraction of $B$ mesons in a sample of 89 million \BB\ events
recorded with  the \babar\ detector at the  \FourS\ resonance.  Events
are selected  by fully reconstructing the  decay of one  $B$ meson and
identifying  a charged  lepton from  the  decay of  the other  $B$
meson. The number of signal events is extracted from the hadronic mass
distribution and is used to determine the ratio of branching fractions
${\BR(\Bxulnu)/\BR(\Bxlnu)}  =  
(2.06 \pm 0.25(stat) \pm 0.23(syst) \pm 0.36(theo))\times 10^{-2}$.
  Using the measured branching fraction for
inclusive semileptonic $B$ decays, we find $\BR(\Bxulnu) = (2.24 \pm
0.27(stat) \pm 0.26(syst) \pm 0.39(theo))\times 10^{-3}$
 and derive the CKM
matrix element $\Vub =
(4.62 \pm 0.28(stat) \pm 0.27(syst)
\pm 0.48 (theo)
)\times 10^{-3} $.
\end{abstract}


\date{\today}%

\maketitle



\ifnum\isprl<1
\section{Introduction}
\label{sec:introduction}
\fi
The element  \Vub of the Cabibbo-Kobayashi-Maskawa quark-mixing matrix~\cite{ckm}
  plays a critical role in testing the consistency of the Standard Model description of \CP\ violation.  In this paper, we present 
a determination of \Vub\ from a measurement of inclusive charmless 
semileptonic decays 
\Bxulnu~\cite{cc}. The analysis uses \FourS\to\BB\ events in which one of the \B\ meson decays hadronically and is fully reconstructed ($B_{reco}$) 
and the semileptonic 
decay of the recoiling \Bb\ meson is identified by the presence of an electron or muon. While this approach 
results in a low overall event selection efficiency, it allows for the determination of the momentum, charge,
 and flavor of the \B\ mesons. 
We use the invariant mass \mX\ of the hadronic system to separate \Bxulnu\ decays from the dominant \Bxclnu\ background, which clusters above the 
\D\ meson mass~\cite{Barger:tz}. 
By ensuring a higher signal purity and  acceptance  than  previously  achieved~\cite{Barate:1998vv}, and 
by measuring   the   fraction    of   charmless    semileptonic   decays
$\rusl={\BR(\Bxulnu)/\BR(\Bxlnu)}$, this analysis leads to  
substantially smaller systematic uncertainties~\cite{duality}.

\ifnum\isprl<1
\section{The \babar\ Dataset and Monte Carlo Simulation}
\label{sec:data}
\fi
\ifnum\isprl<1
\subsection{Data}
\fi

 
The measurement presented here is based on a sample of about 89 million \BB\ pairs
collected  near the \FourS\ resonance by the \babar\ detector~\cite{babarnim} at the
PEP-II  asymmetric-energy  $e^+e^-$ storage ring operating at SLAC.


\ifnum\isprl<1
\subsection{Monte Carlo Simulation}
\label{subsec:montecarlo}
\fi

We use Monte Carlo (MC) simulations of the \babar\ detector based on
\geant~\cite{geant} to optimize selection criteria and to
determine signal efficiencies and background distributions.  
Charmless semileptonic \Bxulnu\ decays  are simulated as a combination
(see Fig.~\ref{fig:mxeff}a)  of resonant  three-body decays  ($X_u =
\pi,  \eta,  \rho,  \omega,   \ldots$)~\cite{ref:isgwtwo}   and  decays   to
nonresonant          hadronic       final       states
$X_u$~\cite{ref:fazioneubert}, for which the hadronization is performed by
\jetset~\cite{ref:jetset}. The motion of the $b$ quark inside the $B$
meson   is  implemented  with   the  shape   function  
parametrization given in Ref.~\cite{ref:fazioneubert}.
%
The simulation of the $\Bxclnu$ background uses an HQET parametrization of form factors for $\Bb\to
D^{*}\ell\nub$~\cite{ref:duboscq}, and models  for $\Bb\to
D \pi \ell\nub, D^* \pi \ell\nub$~\cite{ref:goityroberts}, and  $\Bb\to D
\ell\nub,D^{**}\ell\nub $~\cite{ref:isgwtwo}.

\ifnum\isprl<1
\section{Data Analysis}
\label{sec:analysis}
\fi

\ifnum\isprl<1
\subsection{Selection of Hadronic $B$ Decays, $B \ra \Db Y$}
\fi

To reconstruct a large sample of $B$ mesons, hadronic decays $B_{reco}
\rightarrow  \Db Y^{\pm}, \Db^* Y^{\pm}$  are  selected.  Here, the  system
$Y^{\pm}$ consists of hadrons with a total charge of $\pm 1$, composed
of $n_1\pi^{\pm}\, n_2K^{\pm}\, n_3\KS\,  n_4\piz$, where $n_1 + n_2 \leq
5$,  $n_3  \leq  2$,  and  $n_4  \leq  2$.   We  reconstruct  $D^{*-}\ra
\Dzb\pi^-$; $\Dstarzb \ra
\Dzb\piz, \Dzb\gamma$; $D^-\ra K^+\pi^-\pi^-$, $K^+\pi^-\pi^-\piz$, $\KS\pi^-$,
$\KS\pi^-\piz$, $\KS\pi^-\pi^-\pi^+$; and $\Dzb\ra K^+\pi^-$,
$K^+\pi^-\piz$, $K^+\pi^-\pi^-\pi^+$,  $\KS\pi^+\pi^-$. 
The kinematic consistency of $B_{reco}$ candidates 
is checked with two variables,
the beam energy-substituted mass $\mes = \sqrt{s/4 -
\vec{p}^{\,2}_B}$ and the energy difference 
$\Delta E = E_B - \sqrt{s}/2$. Here $\sqrt{s}$ is the total
energy in the \FourS center of mass frame, and $\vec{p}_B$ and $E_B$
denote the momentum and energy of the $B_{reco}$ candidate in the same
frame.  We require $\Delta E = 0$ within three standard
deviations as measured for each mode.

For each of the 1097 reconstructed $B$ decay modes, the purity
${\cal P}$ is estimated as the fraction of signal events with \mes$>
5.27$\gevcc.  We  only use events for which $\cal P$ exceeds a
decay mode dependent threshold in the range of 9\% to 24\%. 
 In events with more than one
reconstructed $B$ decay, we select the decay mode with the highest purity.  
On average, we reconstruct one $B$ candidate in 0.3\%  (0.5\%) of the 
\BzBzb\ (\BpBm) events. The purity for events
with  a high-momentum lepton is 67\% (see Fig.~\ref{fig:msl}a).

\ifnum\isprl<1
\subsection{Selection of Semileptonic Decays, $\Bxlnu$}
\fi

Semileptonic  decays $\Bxlnu$ of the \Bb\ recoiling against the \breco\
candidate are identified by an electron or muon with a
minimum momentum of $p^* > 1 \gevc$ in the \Bb rest frame.  For
charged \breco\ candidates, we require the charge of the lepton to be
consistent with a prompt semileptonic \Bb\ decay. For
neutral \breco\ candidates, both charge-flavor
combinations are retained and the known average $\Bz$-$\Bzb$ mixing rate is used to extract the prompt lepton yield.
%
%
Electrons are identified~\cite{Aubert:2002uf} with 92\% average efficiency and a
hadron misidentification rate ranging between 0.05\% and 0.1\%.
Muons are identified~\cite{babarnim} with an
efficiency ranging between 60\% ($p^*>1\gevc$) and 75\% ($p^*>2\gevc$)
and hadron misidentification rate between 1\% and 3\%.
Efficiencies and misidentification rates are estimated from
data control  samples.

The hadronic system $X$ in the decay \Bxlnu\ is reconstructed from charged
tracks  and energy depositions
in the calorimeter that are not associated with
the \breco\ candidate or the identified lepton.  Care is taken to eliminate 
fake charged tracks, as well as low-energy beam-generated photons and energy depositions in the calorimeter from charged and
neutral hadrons.  The
neutrino four-momentum $p_{\nu}$ is estimated from the
missing momentum four-vector $p_{miss} = p_{Y(4S)}-p_{\breco} -p_X-p_\ell$, 
where all momenta are measured in the laboratory frame and
$p_{Y(4S)}$ refers to the  \FourS\ meson. 
The mass of the hadronic system is determined by a
kinematic fit that imposes four-momentum conservation, the equality of
the masses of the two $B$ mesons, and forces $p_{\nu}^2 = 0$.  The resulting $\mX$ resolution is 350
\mevcc on average.

To select \Bxulnu\ candidates we require exactly one
charged lepton with $p^* > 1 \gevc$, charge conservation ($Q_{X} +
Q_\ell + Q_{\breco} = 0$), and a missing mass consistent with zero
($\mmiss < 0.5 \gev^2/c^4$).  These criteria suppress the dominant
\Bxclnu\ decays, many of which contain additional neutrinos or
an undetected $\KL$ meson.  We suppress
the $\Bzb\to\Dstarp\ell^-\overline{\nu}$ background by reconstructing
the $\pi^+_s$ from the $\Dstarp\to \Dz\pi_s^+$ decay and the
lepton: since the momentum of the $\pi^+_s$ is almost collinear with the
\Dstarp\ momentum in the laboratory frame, we can approximate the
energy of the \Dstarp\ as $E_{\Dstarp}
\simeq m_{\Dstarp} \cdot E_{\pi_s} /145 \mevcc$ and require for the
neutrino  $  m_{\nu}^2  =  (p_B   -  p_{\Dstarp}  -  p_{\ell})^2  <  -
3\gev^2/c^4$.  We  veto events  with charged or  neutral kaons  in the
recoil  \Bb\ to reduce  the background  from \Bxclnu\  decays.  Charged
kaons are identified \cite{babarnim}
with an efficiency varying between 60\% at the highest and almost 100\% at the
lowest  momenta.  The  pion   misidentification rate is  about  2\%.   The
$\KS\to\pi^+\pi^-$  decays  are reconstructed  with  an efficiency  of
$80\%$ from pairs of oppositely  charged tracks with an invariant mass
between 486  and   510  \mevcc.  The  impact  of  the  event
selection    on   the   $\mX$    distribution   is    illustrated   in
Fig.~\ref{fig:mxeff}b.

\ifnum\isprl<1
   \begin{figure}
    \begin{centering} 
    \hbox{\hskip-1.1cm\epsfig{file=plot0.eps,height=6.cm}
    \epsfig{file=plot3.eps,height=6.cm}
    \epsfig{file=plot5.eps,height=6.cm}}

    \caption{Signal MC $\mX$ distributions: a) generated \mX\
    distributions for the resonant and the nonresonant components of
    the signal model after requiring the
    presence of a lepton with $p^* > 1\gevc$ and  b) measured \mX\ distribution 
after the $p^*$ cut  and
    after all cuts.  \label{fig:mxeff}}

   \end{centering}
   \end{figure} 
\else
   \begin{figure}[t]
    \begin{centering} 
    \hbox{\hskip-0.cm \epsfig{file=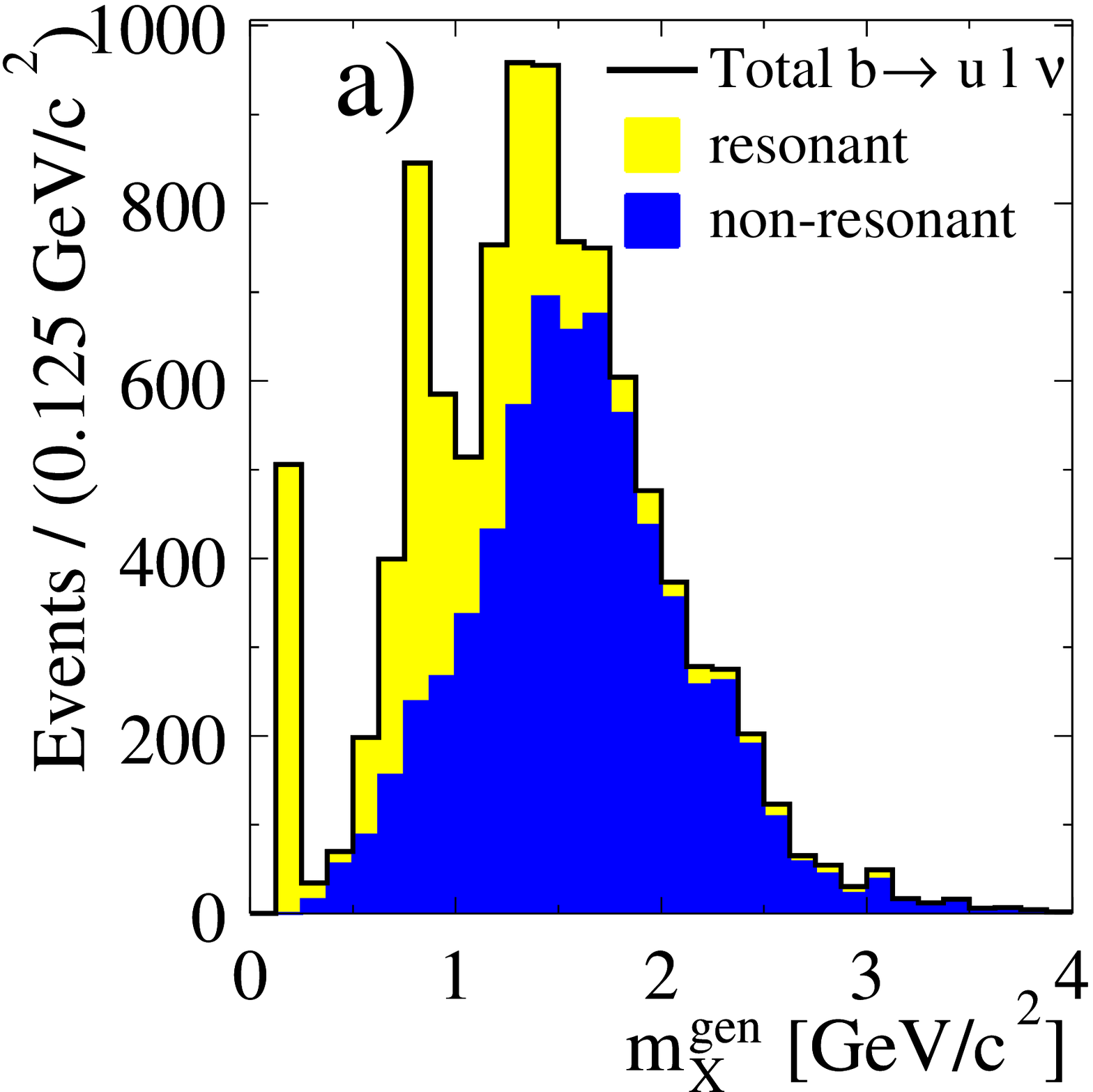,height=4.6cm} \hskip-0.3cm \epsfig{file=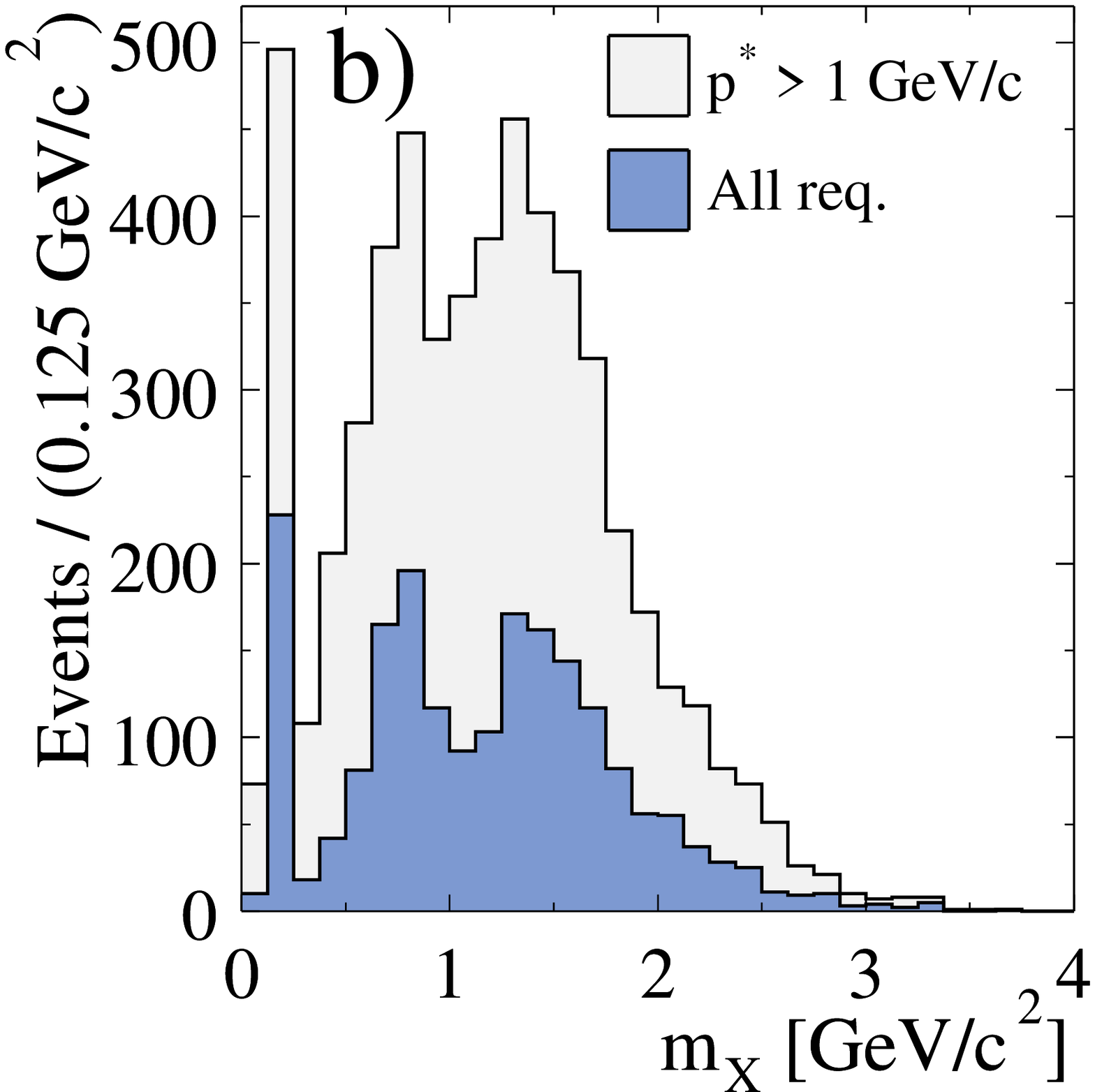,height=4.6cm}}
    \caption{Signal MC $\mX$ distributions with the  requirement of a
    lepton with $p^* > 1\gevc$:  a) generated \mX\ distributions
    for the two components of the signal
    model, and b) measured \mX\ distribution before and after all other requirements.
    \label{fig:mxeff}}
   \end{centering}
   \end{figure} 
\fi

\ifnum\isprl<1
\section{Extraction of Branching Ratios}
\label{sec:fitconcepts}
\fi

We determine \rusl\ from $N_u$, the observed number of $\Bxulnu$ candidates with $\mX<1.55$\gevcc,
 and $N_{sl}$, the number of events with at least one charged lepton: 
\begin{equation*}
\rusl=
\frac{\BR(\Bxulnu)}{\BR(\Bxlnu)}=
\frac{N_u/(\varepsilon_{sel}^u \varepsilon_{\mX}^u)}{N_{sl}} 
\times \frac{\varepsilon_l^{sl} \varepsilon_{reco}^{sl} } {\varepsilon_l^u \varepsilon_{reco}^u }.
\label{eq:vubExtr}
\end{equation*}
Here  $\varepsilon^u_{sel} = (\allepsu\pm0.6)\%$ is the
efficiency for selecting \Bxulnu\ decays once a \Bxlnu\ candidate has been identified, 
 $\varepsilon^u_{\mX} = (\allepsmx\pm0.9)\%$ is the
fraction of signal events with $m_X < 1.55\gevcc$, 
$\varepsilon_l^{sl}/\varepsilon_l^u = 0.887\pm0.008$ corrects for the
difference in the efficiency of the lepton momentum cut for
\Bxlnu\ and
\Bxulnu\ decays, and  $\varepsilon_{reco}^{sl}/\varepsilon_{reco}^u = 1.00 \pm 0.03$
accounts for a possible efficiency difference in the $B_{reco}$
reconstruction in events with \Bxlnu\ and \Bxulnu\ decays.

 We derive $N_{sl}$  from a fit to the \mes 
distribution shown in Fig.~\ref{fig:msl}a.
The fit uses an empirical description~\cite{argusf} of the combinatorial 
background from continuum and \BB\ events, together with a narrow
 signal~\cite{cry} peaked at the $B$ meson mass. The small tail accounts
for energy losses in the reconstruction of \piz\ mesons. 
   The  residual  background  in  $N_{sl}$  from  misidentified
leptons and semileptonic charm decays amounts to $6.8\%$ and
is subtracted.

\ifnum\isprl<1
   \begin{figure} \begin{centering}
    \epsfig{file=mes-nsl.eps,height=8.cm}
    \epsfig{file=mes-ball-lomx.eps,height=8.cm}

    \caption{a)  Fit to  the  $\mes$ distributions  for the  inclusive
    lepton  sample with  $p^* >  1\gevc$ in  the recoil  of  a \breco\
    candidate, and  b) for selected  events with $\mX <  1.55 \gevcc$.
    The  arrow  indicates the  cut  on  \mes\  for signal  candidates.
    \label{fig:msl}} \end{centering} \end{figure}
\else
   \begin{figure}[t]
    \begin{centering}
    \hbox{\hskip-0.cm \epsfig{file=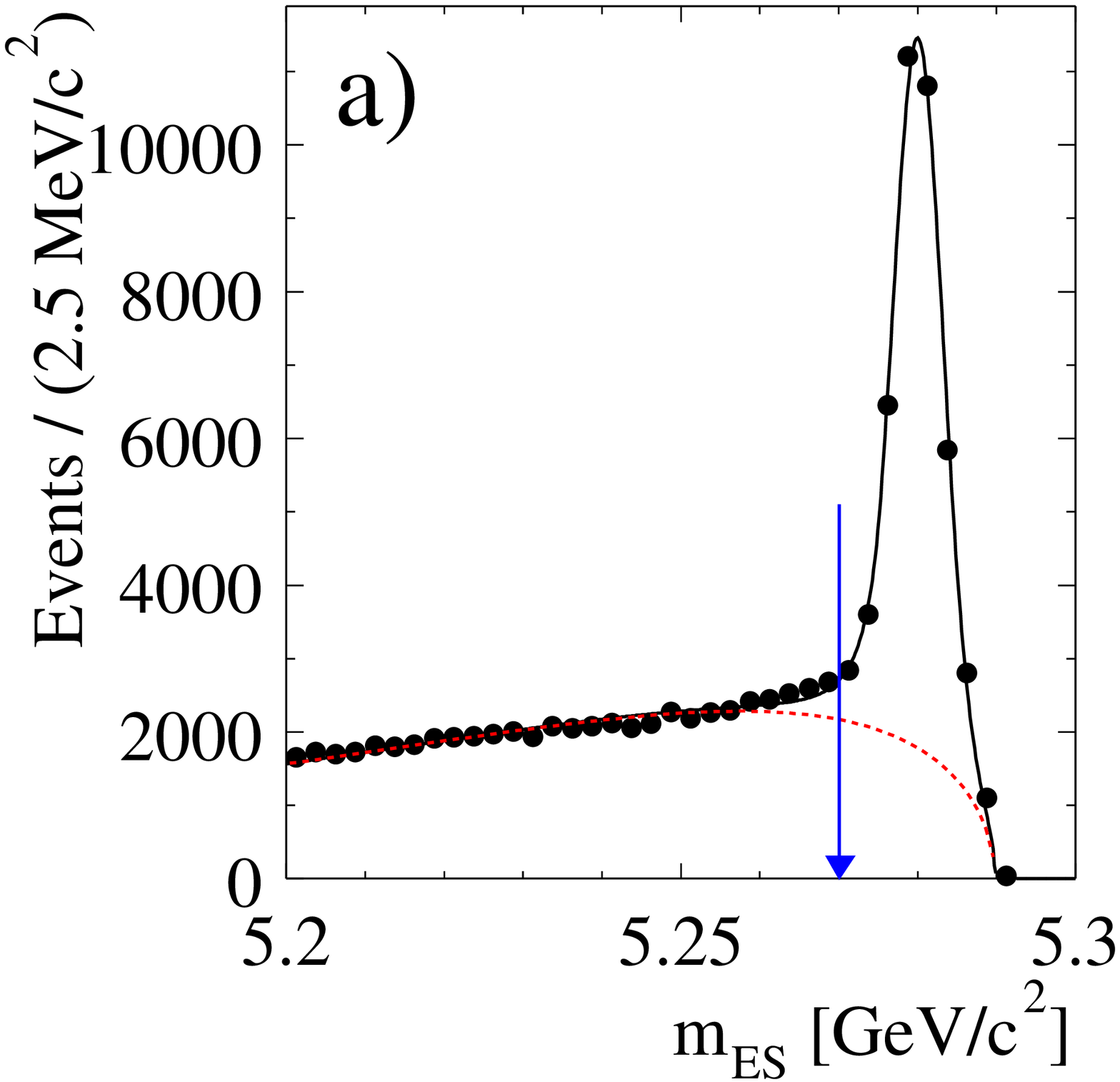,height=4.6cm} \hskip-0.4cm \epsfig{file=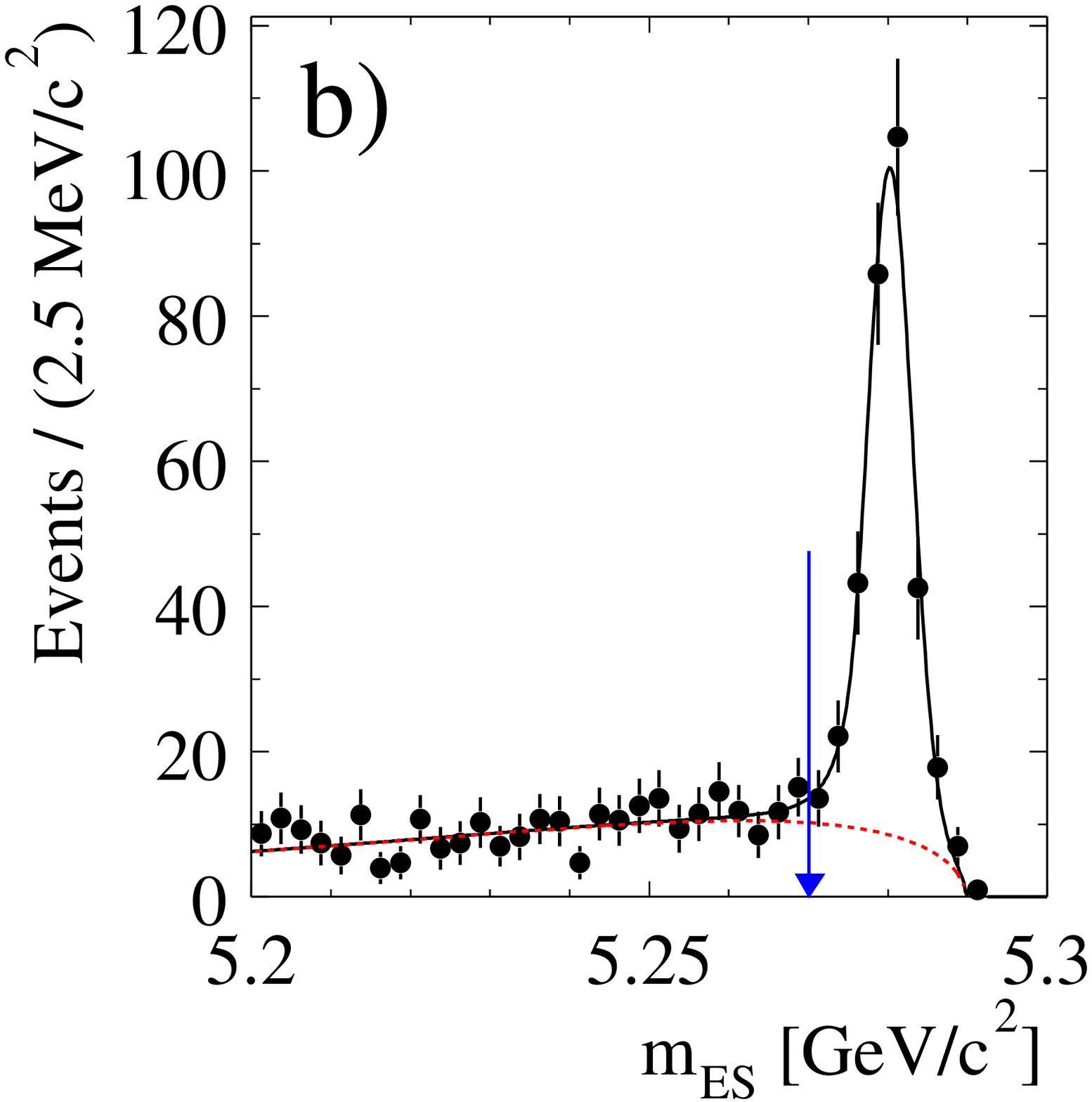,height=4.6cm}}

    \caption{Fit to the $\mes$ distributions for a) the sample with a $p^*>1\gevc$ lepton 
 and b) the sample after all requirements and  with $\mX < 1.55 \gevcc$. 
 The arrow    indicates the lower limit of the signal region.
    \label{fig:msl}}

   \end{centering}
   \end{figure}
\fi
We extract $N_u$ from   the $\mX$ distribution by a minimum $\chi^2$
fit to the sum of three contributions: the signal, the background
$N_{c}$ from
\Bxclnu, and a background of $<1\%$ from other sources (misidentified
leptons, secondary $\tau$ and charm decays). 
 In each bin of the
\mX\ distribution, the combinatorial $B_{reco}$ background for $\mes>5.27$ is
subtracted on the basis of a fit to the $\mes$ distribution (Fig.~\ref{fig:msl}b). 
Fig.~\ref{fig:fitdata}a shows the fitted $\mX$ distribution.
To minimize the model dependence,
the first bin is extended to $\mX < 1.55\gevcc$.
 The fit reproduces the data well with $\chi^2/dof=7.6/6$.
Fig.~\ref{fig:fitdata}b shows the $\mX$ distribution
after background subtraction
with finer binning.
 Table~\ref{breakdown} summarizes the results of fits with different requirements on \mX,
 for electrons and muons, for neutral and charged \breco\ candidates, 
and 
for different ranges of the $B_{reco}$ purity $\cal P$.
The results are all consistent within the uncorrelated statistical errors.

\ifnum\isprl<1
   \begin{figure}
    \begin{centering}
     \hbox{\hskip-1cm\epsfig{file=fitresult1.eps,height=6.0cm} \epsfig{file=fitresult0.eps,height=6.0cm} \epsfig{file=fitresult2.eps,height=6.0cm}}

     \caption{The $\mX$ distribution in \Bxlnu\ decays. a) Selected
     data (points) with a $\chi^2$ fit. b) Background subtracted
     data and signal MC.  \label{fig:fitdata}}

   \end{centering}
   \end{figure} 
\else
\begin{figure}[t]
    \begin{centering}
     \hbox{\hskip -0.cm \epsfig{file=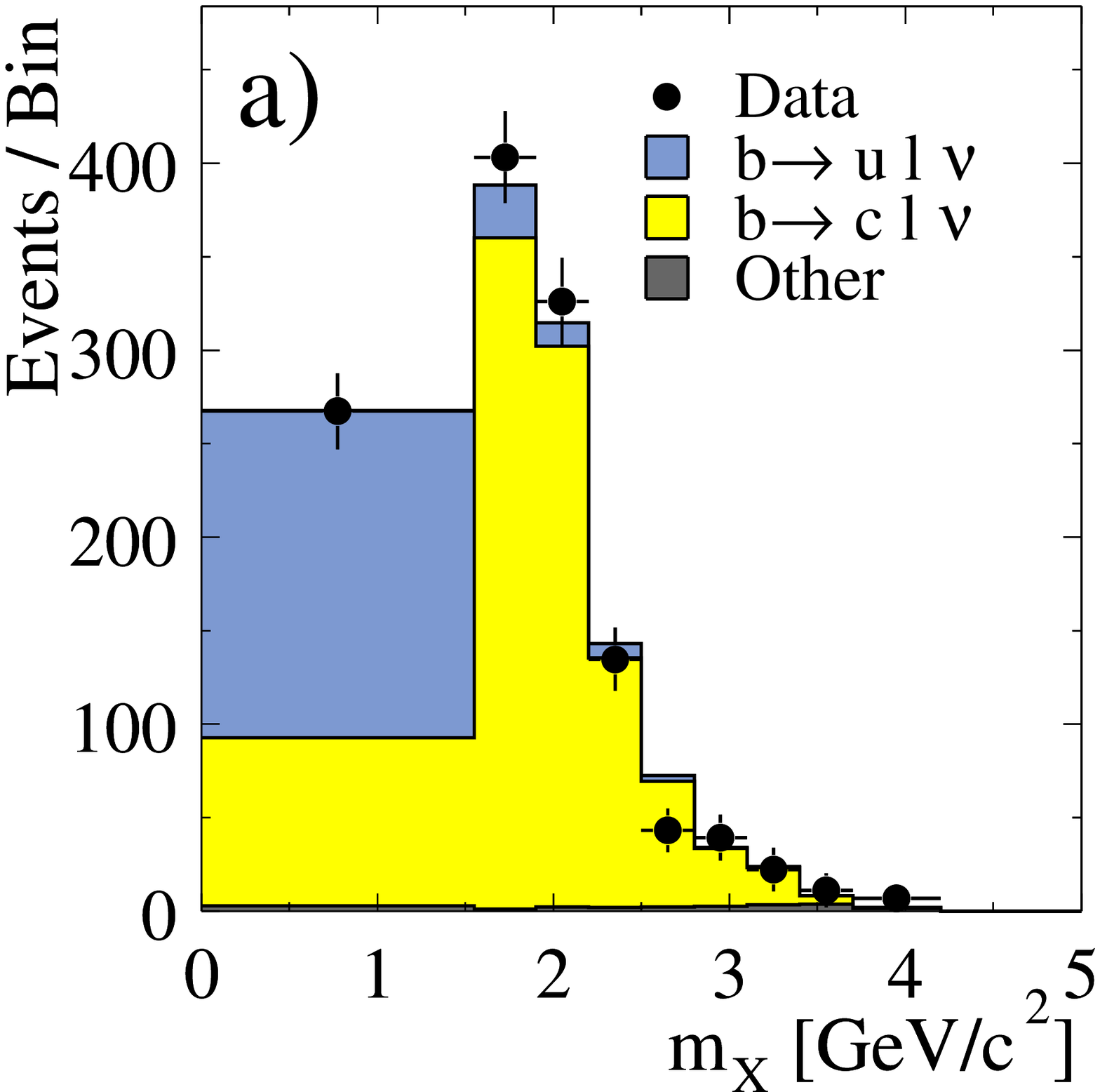,height=4.6cm} \hskip-0.3cm \epsfig{file=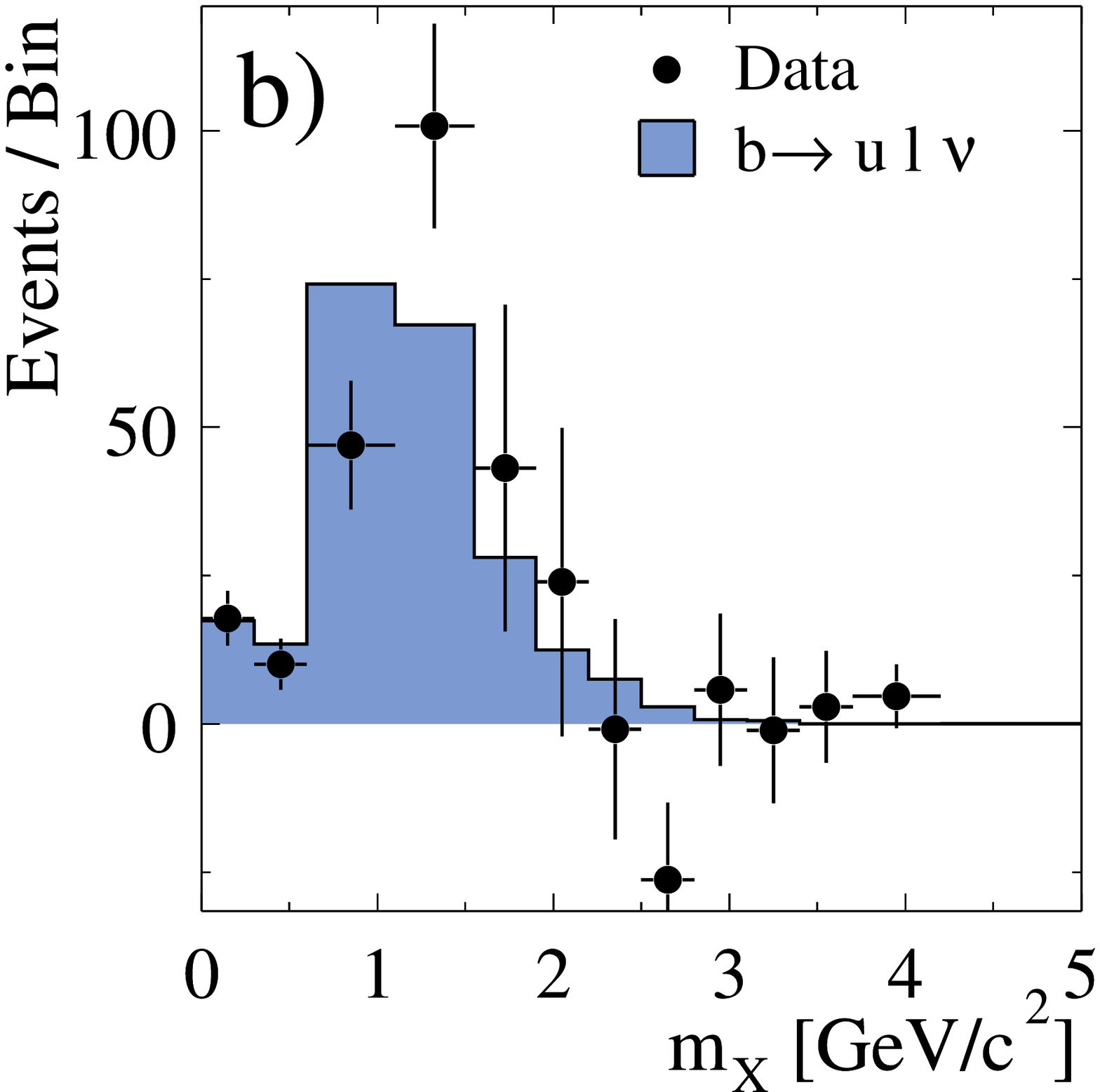,height=4.6cm} }

     \caption{The $\mX$ distribution for \Bxlnu\ candidates: a)
     data (points) and fit components, and b) 
     data and signal MC after subtraction of the $b\to c\ell\nu$ and the ``other'' backgrounds.  \label{fig:fitdata}}

   \end{centering}
   \end{figure} 
\fi

\begin{table}
\begin{center}
\caption{Fit results for data subsamples.}
\scriptsize
\vspace{0.03in}
\begin{tabular}{|lcccc|} \hline 
Sample  &$N_{sl}$&$N_u$& $N_{c}$ &$\rusl (\%)$\\ 
\hline
$\mX<1.55\gevcc$     &$\allslsub$      &$\allnu$     &$\allbgc$     &$\allbrbr $\\
$\mX<1.40\gevcc$     &$\lomxslsub$     &$\lomxnu$    &$\lomxbgc$    &$\lomxbrbr $\\
$\mX<1.70\gevcc$     &$\himxslsub$     &$\himxnu$    &$\himxbgc$    &$\himxbrbr $\\
\hline                                                              
neutral $B_{reco}$             &$\bneslsub$      &$\bnenu$     &$\bnebgc$     &$\bnebrbr $\\
charged $B_{reco}$         &$\bchslsub$      &$\bchnu$     &$\bchbgc$     &$\bchbrbr$\\
\hline                                                              
Electrons            &$\eleslsub$      &$\elenu$     &$\elebgc$     &$\elebrbr$\\
Muons                &$\muslsub$       &$\munu$      &$\mubgc$      &$\mubrbr$\\
\hline                                                              
${\cal P}>80\%$      &$\sboneslsub$    &$\sbonenu$   &$\sbonebgc$   &$\sbonebrbr$  \\
$50\%<{\cal P}<80\%$ &$\sbtwoslsub$    &$\sbtwonu$   &$\sbtwobgc$   &$\sbtwobrbr$  \\
${\cal P}<50\%$      &$\sbthreeslsub$  &$\sbthreenu$ &$\sbthreebgc$ &$\sbthreebrbr$\\
\hline
\end{tabular}
\label{breakdown}
\end{center}
\end{table}

We have performed extensive studies to determine the systematic
uncertainties on \rusl. 
To establish that the
background from \Bxclnu\ events is adequately simulated we use previously excluded events with charged or neutral kaons as a control sample.
The relative systematic error due to uncertainties in the detection of photons is estimated to be 4.7\% by varying the corrections applied 
to the MC simulation to match the data control  samples.  An additional error of 1.0\% is ascribed to the uncertainty in the simulation of 
showers generated by \KL\ interactions; it is equal to the shift caused by the removal of the \KL\ energy depositions in the MC simulation.
 
An error of 1.0\% is attributed to the uncertainty in the track-finding efficiency.
The error due to identification of electrons, muons, and kaons is estimated to be 1.0\%, 1.0\%, and 2.3\%, respectively,  by
varying identification efficiency by $\pm
2\%$, $\pm3\%$, and $\pm2\%$ for $e^{\pm}, \mu^{\pm}$, and $K^{\pm}$,  and  the misidentification rates by $\pm15\%$ for all particle types.

The uncertainty in
the $B_{reco}$ combinatorial background subtraction contributes 3.8\%. It is estimated by
changing the empirical $\mes$ signal function to a Gaussian distribution and by varying the parameters within one standard deviation of the default values. 
The limited statistics of the simulated event samples adds an uncertainty of 4.5\%.  The choice of bins for $\mX >
1.55\gevcc$ impacts the fit result at a level of 1.2\%. 
All the above mentioned experimental errors add up to 8.7\%.

The uncertainties in the background modeling due to branching fraction measurements for $B\to D\ell\nu, D^*\ell\nu,...$ and for inclusive and exclusive $D$ meson decays ~\cite{pdg2002} contribute 4.4\%.

The error due to the hadronization in the
\Bxulnu\ final state is estimated to be 3.0\% by measuring $\rusl$ as a function of the charged and neutral particle multiplicities and performing the fit
with only the nonresonant part of the signal model.
We assign an additional 2.8\% error to account for the uncertainties in the inclusive and exclusive 
branching fractions for charmless semileptonic $B$ decays ~\cite{pdg2002}, plus 3.7\% 
for the veto on strange particles.  Here, we assume a 100\% uncertainty in the  \ssbar\ contents for the resonant and 30\%
for the nonresonant component ~\cite{Althoff:1984iz}.  These three uncertainties contribute a combined error of 5.5\%.

The  efficiencies $\varepsilon_{sel}^{u}$  and $\varepsilon_{\mX}^{u}$
are sensitive to the detailed  modeling of the \Bxulnu\ decays. 
We assess these
uncertainties by  varying the  nonperturbative parameters in  the model
\cite{ref:fazioneubert} within  their errors,  $\lbar =  0.48  \pm 0.12 
\gev$ and  $\lone =  -0.30\pm0.11  \gev^2$,
obtained from the results in Ref.~\cite{ref:cleomxhad} by removing
terms proportional to $1/m_b^3$ and $\alpha_s^2$ from the relation
between the measured observables and \lbar\ and \lone.
%
Taking  into
account the correlation of $-0.8$ between \lbar\ and \lone, we arrive at a theoretical error of 17.5\%.  

In summary, we obtain
\begin{equation*}
\rusl = (\allbrbrVal \pm \allbrbrEcstat
\pm \allbrbrEsyst
\pm\allbrbrEthHi)\times 10^{-2},
\end{equation*}
where the errors are statistical, systematic (experimental plus signal and background modeling), and theoretical, respectively. 
Taking into account common errors we compute
the  double ratio 
$\frac{\BR(\Bzxulnu)}{\BR(\Bzxlnu)}\frac{\BR(\Bpxlnu)}{\BR(\Bpxulnu)}
=0.72 \pm 0.18 (stat) \pm 0.19 (syst)$.  Combining the ratio
\rusl\ with the measured inclusive semileptonic branching fraction 
$\BR(\Bxlnu) = (10.87 \pm 0.18(stat) \pm
0.30(syst))\%$~\cite{Aubert:2002uf}, we have
\begin{equation*}
\BR(\Bxulnu) = 
(\allbr
\pm\allbrE
\pm\allbrEsyst
\pm\allbrEthHi)\times10^{-3}.
\end{equation*}

We combine this result with 
the average $B$ lifetime of $\tau_B = 1.608 \pm 0.012 \ps$~\cite{pdg2002,b0bch} and obtain~\cite{Uraltsev:1999rr}
\begin{equation*}
\Vub = (\allvub
\pm\allvubE
\pm\allvubEsyst
\pm\allvubEthHi
\pm\allvubEtheo)\times 10^{-3}.
\end{equation*}
The first error is  statistical, the second  systematic, the third gives the theoretical uncertainty in the signal efficiency and 
the extrapolation of \rusl\ to the full $\mX$ range, 
and the fourth combines the perturbative and
nonperturbative uncertainties in the extraction of \Vub\ from the
total decay rate.


This result is consistent with previous inclusive
measurements~\cite{Barate:1998vv},
but has a smaller systematic error,
primarily due to larger phase-space acceptance and much higher sample purity.
In the future, improved understanding of the signal composition and
charm background will significantly reduce the experimental errors,
and this, together with independent measurements of $ b \to s$
transitions and semileptonic $B$ decays, is expected to constrain the
theoretical uncertainties.

We are grateful for the excellent luminosity and machine conditions
provided by our \pep2\ colleagues, 
and for the substantial dedicated effort from
the computing organizations that support \babar.
The collaborating institutions wish to thank 
SLAC for its support and kind hospitality. 
This work is supported by
DOE
and NSF (USA),
NSERC (Canada),
IHEP (China),
CEA and
CNRS-IN2P3
(France),
BMBF and DFG
(Germany),
INFN (Italy),
FOM (The Netherlands),
NFR (Norway),
MIST (Russia), and
PPARC (United Kingdom). 
Individuals have received support from the 
A.~P.~Sloan Foundation, 
Research Corporation,
and Alexander von Humboldt Foundation.

\label{sec:references}


\begin{thebibliography}{02}
\bibitem{cc}
Charge-conjugation  is  implied  throughout the Letter.
\bibitem{duality} The results of this analysis depend on the
assumption of   parton-hadron duality to which no error is assigned.
\bibitem{b0bch} The impact of the uncertainty of the fraction of neutral and charged $B$ mesons is negligible.
\end{thebibliography}
\end{document}